**Title:** Analytical Solution and Parameter Estimation for Heat of Wetting and Vapor Adsorption During Spontaneous Imbibition in Tuff

**Authors:** Forest T. Good[1,4], Kristopher L. Kuhlman[1], Tara C. LaForce[2], Matthew J. Paul[1] & Jason E. Heath[3]

**Affiliations:** Sandia National Laboratories, [1]Nuclear Waste Disposal Research & Analysis Department, [2]Applied Systems Analysis & Research Department, [3]Geomechanics Department. [4]New Mexico Institute of Mining and Technology

**ABSTRACT**

An analytical expression is derived for the thermal response observed during spontaneous imbibition of water into a dry core of zeolitic tuff. Sample tortuosity, thermal conductivity, and thermal source strength are estimated from fitting an analytical solution to temperature observations during a single laboratory test. The closed-form analytical solution is derived using Green's functions for heat conduction in the limit of "slow" water movement; that is, when advection of thermal energy with the wetting front is negligible. The solution has four free fitting parameters and is efficient for parameter estimation. Laboratory imbibition data used to constrain the model include a time series of the mass of water imbibed, visual location of the wetting front through time, and temperature time series at six locations. The thermal front reached the end of the core hours before the visible wetting front. Thus, the predominant form of heating during imbibition in this zeolitic tuff is due to vapor adsorption in dry zeolitic rock ahead of the wetting front. The separation of the wetting front and thermal front in this zeolitic tuff is significant, compared to wetting front behavior of most materials reported in the literature. This work is the first interpretation of a thermal imbibition response to estimate transport (tortuosity) and thermal properties (including thermal conductivity) from a single laboratory test.

**Research Highlights:**

1. Zeolitic tuff core shows thermal pulse from vapor transport is much faster than wetting front
2. Data from heat of wetting spontaneous imbibition test used to estimate tortuosity to water vapor diffusion
3. Analytical solution for heat conduction used to estimate thermal conductivity and source strength

**INTRODUCTION**

Imbibition of water into porous media is important in many applications in a range of rocks, soils, or engineered materials, including water migration through the vadose zone associated with agricultural, groundwater-hydrology, and engineering applications in man-made porous media (Hillel, 2004). Infiltration of water is important in the vadose zone, sometimes perching on obstructions to vertical flow (Kwicklis et al., 2019). There is a need to parameterize two-phase fluid flow and heat conduction properties of rock samples during site characterization efforts and to improve understanding of fundamental physics governing the fate of water in deep vadose zones, for example, associated with underground nuclear explosions (Bourret et al., 2019; Heath et al., 2021) or radioactive waste disposal (Wang & Bodvarsson, 2003; Rechard et al., 2014).

Imbibition of water into a dry porous medium releases heat associated with both the adsorption of water vapor and the wetting of pores with liquid water. Condensation of vapor to liquid water releases energy and adsorption of water vapor or bulk liquid water to solid surfaces releases energy. The heat associated with the movement of liquid water into a porous medium is called heat of wetting (Edlefsen & Anderson, 1943), while the heat associated with water vapor adsorption is called sorptive heating (Murali et al., 2020) or the heat of adsorption (Aslannejad et al., 2017). Heat released during wetting a porous medium is well known and has historically been called the Pouillet Effect (Adam, 1941; Pouillet, 1822), but the first mention in the scientific literature is likely Leslie (1802), who stated "I have fometimes produced a heat of ten degrees by moiftening faw-duft which had been parched before the fire [sic]." Possible practical applications of the phenomenon include laboratory material property testing, field observation of water fronts in multiphase oil and gas systems (temperature changes are simpler to measure than saturation changes and may provide an early warning of encroaching water to production wells), or even medical diagnostics (Aslannejad et al., 2017).

The energy released by the heat of wetting or the heat of adsorption is proportional to the media's specific surface area; larger temperature rises are seen in clays and zeolites, while smaller temperature rises are seen in sands or glass beads (Anderson & Linville, 1962). The theory governing this thermal-hydrological process has been discussed in soil science (Edlefsen & Anderson, 1943; Ten Berge & Bolt; 1988), hydrology (de Vries, 1958), paper microfluidics (Murali et al., 2020), and grain storage (Thorpe & Whitaker, 1992). Experimental data have been presented in the literature for imbibition into or immersion of paper (Foss et al., 2003; Aslannejad et al., 2017; Terzis et al., 2018; Murali et al., 2020), textiles (Bright et al., 1953), glass beads (Perrier & Prakash, 1977), starch (Janert, 1934), wood (Hearmon & Burcham, 1955), loamy and clay soils (Anderson & Linville, 1962; Prunty & Bell, 2005), zeolite (Żołądek-Nowak et al., 2012), coal (Nordon & Bainbridge, 1983), and clay (Anderson & Linville, 1960). The heat of wetting effect is used in immersion calorimetry to estimate rock wettability in oil-water systems (Korobkov et al., 2016) and colloidal content in soils (Anderson, 1924).

Although there are many observations of heat released during adsorption of water or vapor, there is not universal agreement in the soil science literature as to the relative importance of these possible two energy sources, with most studies stating either one or the other is much more important. Earlier work in the soil science literature considered the greater importance of vapor transport (Anderson & Linville, 1960; Perrier & Prakash, 1977). It has been observed that less heat is released when imbibing into wetter materials, but some heat is still released during imbibition into pre-wetted media (Anderson et al., 1963; Nordon & Bainbridge, 1983). Polar fluids (i.e., water) tend to release more heat than non-polar fluids, especially in clays or organic-rich soils (Janert, 1934). Whether heat of wetting or vapor adsorption, Adam (1941) states the thermal energy released is associated with the portion of the liquid phase most strongly interacting with the solid phase. In contrast, condensation of vapor into the liquid phase would not require interaction with a solid phase. More recently, Prunty & Bell (2005) acknowledged that adsorption associated with vapor transport could result in an observed thermal signal but believed it would be of minimal importance compared to the heat of wetting. Aslannejad et al. (2017) went as far as to claim the temperature rise due to adsorption of water vapor "does not hold," since "vapor can diffuse relatively fast in the pore space and should therefore condensate everywhere." Nordon & Bainbridge (1983) thermodynamically treat vapor adsorption as consisting of condensation of vapor to liquid and adsorption of the liquid phase.

Several numerical models have implemented various aspects of both these mechanisms (Milly & Eagleson, 1980; Benjamin et al., 1990; Prunty, 2002; Foss et al., 2003; Prunty & Bell, 2005; Murali et al., 2020), but no analytical solutions have yet been developed to describe the propagation and expected time evolution of a thermal signal observed in a sample during imbibition (spontaneous or forced). Analytical solutions are more efficient, readily allowing parameter estimation, and their development (i.e., dimensional analysis) often leads to further insight. Both analytical and numerical solutions are available to describe the imbibition of water into a porous medium (see references in Kuhlman et al., 2022b), depending on the configuration (i.e., perpendicular to, with, or against gravity), the geometry (i.e., imbibition from one end, or imbibition from all sides during submergence), and the driving force (i.e., free or forced imbibition). Herein is presented an analytical expression and solution for a one-dimensional thermal response during spontaneous imbibition.

The balance of conduction and advection of energy within a sample is characterized by the Péclet number. In the limit of high Péclet number, advection is stronger than conduction of heat. The analytical solution is fit to data from a lower-permeability tuff core associated with small Péclet number, so only heat conduction is considered.

Prunty and Bell (2005) present an infiltration transient temperature test, where water is applied at the top of a soil sample (i.e., flow with gravity). A similar test is presented here for spontaneous imbibition up a cylindrical tuff core sample (i.e., against gravity), with temperature sensors attached to the side of the core. The results of the test presented here show the importance of vapor transport far ahead of the wetting front. The following sections present the mass transport in terms of vapor- and wetting-front sorptivities used by the analytical solution, and the thermal transport model is developed using the Green's function solution. The solution is then fit to an imbibition experiment conducted in zeolitic tuff, and a summary concludes with thoughts about the usefulness and limitations of the analytical solution.

## SOLUTION DEVELOPMENT

### Mass Transport Model

There are two possible fronts associated with imbibition into a dry porous medium, as discussed in the introduction. One possibility is the wetting front, and the second is vapor diffusion ahead of the wetting front.

For a temperature perturbation associated with the heat of wetting (i.e., vapor adsorption is not significant ahead of the wetting front), the height of the wetting front for spontaneous imbibition against gravity is proportional to the square root of time (Green & Ampt, 1911; Washburn, 1921). The sorptivity of the wetting front ($\sigma_w$) [m/$\sqrt{s}$] is the slope of the wetting front position against square-root time (Philip, 1957), and the position of the wetting front is given by $\sigma_w\sqrt{t}$ (Figure 1). The wetting front sorptivity itself has been related via regression to the permeability and porosity of rocks and soils (Tokunaga, 2020).

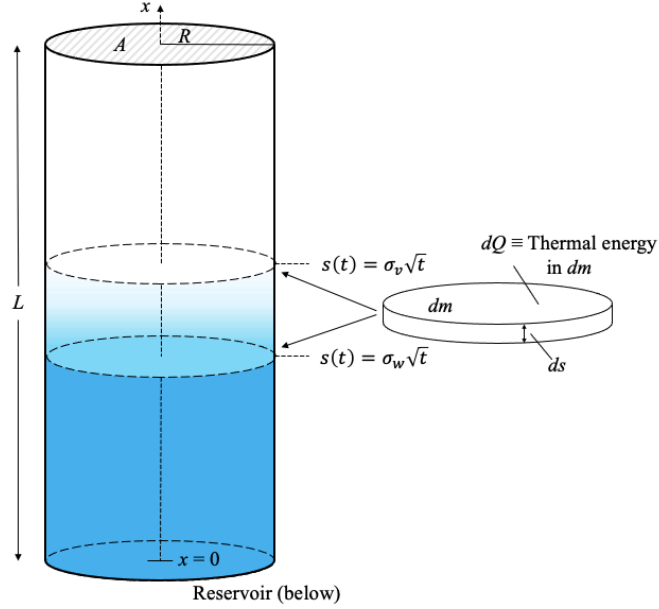

**Figure 1. Conceptual model of mass and energy transport in a cylindrical core.**

For a temperature perturbation associated with vapor adsorption (i.e., the wetting front is moving slowly), the sorptivity of the vapor front ($\sigma_v$) [m/√s] ahead of a nearly static wetting front is proportional to the effective diffusivity of water vapor in the medium. The effective diffusivity is $\mathfrak{D}_{eff} = D_{wv,air}\phi\bar{\tau}(1 - S_l)$ [m²/s], where $D_{wv,air} = 2.6 \times 10^{-5}$ m²/s (Lee & Wilke,1954) is the diffusivity of water vapor in air at ambient conditions, $\bar{\tau}$ is the medium tortuosity to water vapor diffusion in air, and $S_l$ is the liquid saturation. The diffusional tortuosity given here and a classical geometrical tortuosity (the shortest path length divided by true path length squared) are related but are not necessarily equal (Clennel, 1997).

Diffusion of the leading edge (here assumed 50% relative humidity – RH; similar to the minimum seen by Perrier & Prakash, 1977) of the vapor front away from a stationary wetting front (100% RH) into a dry medium (assumed zero RH) is $0.5 = \text{erfc}[z/\sqrt{4\mathfrak{D}_{eff}t}]$ (Carslaw & Jaeger,1959; §2.5, Eqn 3), where $z$ is distance from the wetting front and $t$ is time since imbibition began. This expression is solved for the position of the 50% RH contour, $z = \text{erfc}^{-1}(0.5)\sqrt{4\mathfrak{D}_{eff}t}$, where $\text{erfc}^{-1}(0.5) = 0.477$, such that $\sigma_v = 0.954\sqrt{\mathfrak{D}_{eff}}$. The effective diffusion coefficient is then related back to the diffusivity in air as $\sqrt{\mathfrak{D}_{eff}} = \sqrt{D_{wv,air}\phi\bar{\tau}}$ (for a dry porous medium, $S_l = 0$).

**Energy Transport Model**

Consider the heat conduction problem in a one-dimensional cylindrical domain $0 \leq x \leq L$ with a moving heat source associated with the heating rate given by $g(x, t)$

$$\frac{\partial}{\partial x}\left(k\frac{\partial T}{\partial x}\right) + g(x,t) = \frac{\partial}{\partial t}(\rho c_p T) + h\frac{P}{A}T$$

$$-k\frac{\partial T}{\partial x}\bigg|_{x=0} + h_1 T = 0 \qquad (1)$$

$$k\frac{\partial T}{\partial x}\bigg|_{x=L} + h_2 T = 0$$

$$T(x,0) = 0$$

where $T$ is the change in temperature above the initial background [K], $k$ is the bulk porous medium thermal conductivity [W/(m · K)], $\alpha = k/\rho c_p$ is the bulk thermal diffusivity [m/s$^2$], $\rho$ is the bulk density [kg/m$^3$], $c_p$ is the bulk specific heat capacity [J/(kg · K)], $P$ is the perimeter of the cylindrical core [m], $A$ is the cross-sectional area to flow [m$^2$], $h$ is the heat transfer coefficient representing heat loss to the curved sides of the cylindrical core [W/(m$^2$ · K)], and $h_i$ ($i = 1,2$) are similar heat loss coefficients for the ends of the core. In this model, thermal conductivity, bulk density, and specific heat capacity are assumed to be constant over the range of saturations and temperatures encountered.

This imbibition problem is like the Stefan phase-change problem (Özişik, 1993) since an interface is releasing latent heat as it moves through a solid. In the Stefan problem, the temperature and pressure are fixed at the phase change condition, but the position of the phase change boundary is unknown. In this problem, the position of the interface is known (from the hydrologic problem), the temperature at the interface is unknown, and the source of energy released at the interface is proportional to the medium properties (e.g., specific surface area or organic content).

A heating rate associated with the release of energy by latent heat is assumed as

$$g(x,t) = W\delta[x - s(t)], \qquad (2)$$

where $\delta$ is the Dirac delta function which is zero everywhere except at $x = s(t)$ (the position of the thermal front [m]). Although physically motivated, the assignment of the heat of wetting or vapor adsorption to a moving point in space is a simplification taken to make the problem mathematically tractable. Heating occurs over a narrow band near the leading edge of the wetting or vapor adsorption front – as evidenced by thermal images of imbibition into paper (Aslannejad et al., 2017; Murali et al., 2020). In the limit of a narrow pore-size distribution, this would approach a Dirac delta function. The heating rate has magnitude $W$ [W/m$^2$], which is found from the amount of heat released due to water or vapor adsorption in a differential slab of length $ds$ in the $x$ direction (Figure 1). This differential heat $dQ$ [J] is associated with $dQ = \eta dm$, where $\eta$ is the heat of adsorption per unit mass of the porous media [J/kg] and $dm$ is the mass of the differential slab [kg]. In this cylindrical geometry, the rate water or vapor adsorbs to solid mass is proportional to the wetting or vapor front speed, $\frac{dm}{dt} = \rho A \frac{ds}{dt}$, which gives $W = \left(\frac{dQ}{dt}\right)/A = \rho\eta\frac{ds}{dt}$. This is the same expression for the heat produced at the phase change interface in the Stefan problem (Özişik, 1993), but here the energy is associated with a moving Dirac delta heat source, rather than the interface condition at the unknown boundary location. This results in a heat source given by

$$g(x,t) = \rho\eta \frac{ds}{dt}\delta[x - s(t)]. \tag{3}$$

The thermal front position, $s(t)$, is the product of the generic sorptivity, $\sigma$ [m/√s] (either $\sigma_v$ or $\sigma_w$), and a measure of elapsed square root time (Green & Ampt, 1911), $s(t) = \sigma\sqrt{t}$ (Figure 1). Given this, the front speed is $ds/dt = \sigma/(2\sqrt{t})$. This term can represent the position of the wetting front (i.e., heat of wetting, where $\sigma = \sigma_w$; Philip, 1957) or vapor adsorption front ($\sigma = \sigma_v$) and results in a heating rate of

$$g(x,t) = \rho\eta \frac{\sigma}{2\sqrt{t}}\delta(x - \sigma\sqrt{t}). \tag{4}$$

**Solution using Green's Functions**

The one-dimensional transient heat conduction equation on the region $0 \leq x \leq L$ with a moving heat source ( 4 ) and heat loss to the cylindrical core boundary becomes

$$\frac{\partial^2 T}{\partial x^2} + \frac{1}{k}\rho\eta \frac{\sigma}{2\sqrt{t}}\delta(x - \sigma\sqrt{t}) = \frac{1}{\alpha}\frac{\partial T}{\partial t} + \frac{Ph}{Ak}T; \quad 0 \leq x \leq L, \quad 0 \leq t \leq (L/\sigma)^2. \tag{5}$$

The solution is valid in the limit where the conduction of heat away from the front is faster than the migration of the front since there is no thermal advection term (i.e., small Péclet number). This approximation should be valid in situations where advection of energy with the liquid through the pore space is low, while thermal conductivity through the bulk (i.e., both solid and porous fractions) is high.

At the boundary where imbibition begins ($x = 0$), the Robin boundary condition can be simplified to a Dirichlet boundary condition if it is assumed that the temperature is held constant. This is because the boundary is in constant contact with the reservoir and maintains the same temperature as the water in the reservoir. The boundary and initial conditions can then be formulated as

$$\begin{array}{c} T(0,t) = 0 \\ k\frac{\partial T}{\partial x}\bigg|_{x=L} + h_2 T(L,t) = 0 \\ T(x,0) = 0. \end{array} \tag{6}$$

The problem can now be re-formulated in terms of the dimensionless variables

$$\chi = \frac{x}{L}; \quad \tau = \frac{\alpha t}{L^2}; \quad \theta = \frac{kT}{\rho\alpha\eta} = \frac{c_p T}{\eta}. \tag{7}$$

The transformations for $x$ and $t$ are typical dimensionless transformations for heat conduction problems, while the transformation of $T$ uses a reference temperature $T_r = \eta/c_p$. This is the rise of temperature of an infinitesimal slice of rock due to the vapor or water adsorption releasing energy $\eta$ in the region, useful when the initial conditions and both boundary conditions are zero. These transformations reduce the governing partial differential equation to

$$\frac{\partial^2 \theta}{\partial \chi^2} + \frac{v}{2\sqrt{\tau}}\delta(\chi - v\sqrt{\tau}) = \frac{\partial \theta}{\partial \tau} + C\theta; \quad 0 < \chi < 1, \quad 0 < \tau < \frac{1}{v^2} \tag{8}$$

$$\theta(0,\tau) = 0$$

$$\left.\frac{\partial \theta}{\partial \chi}\right|_{\chi=1} + H_2 \theta(1, \tau) = 0$$

$$\theta(\chi, 0) = 0,$$

where $v = \frac{\sigma}{\sqrt{\alpha}}$ is the ratio of mass advection and thermal diffusion (the inverse square root of the Lewis number), $H_2 = \frac{h_2 L}{k}$ is the Biot number, and $C = \frac{PhL^2}{Ak}$ is proportional to the Biot number (relating heat loss at a boundary to heat conduction). The dimensionless parameter $H_2$ relates heat loss out the end of the core to conduction, while $C$ relates heat loss out the sides of the core to conduction. The dimensionless source term becomes $g^* = \frac{v}{2\sqrt{\tau}} \delta(\chi - v\sqrt{\tau})$.

This dimensionless problem has the Green's function solution (Özişik, 1993) of

$$\theta(\chi, \tau) = \int_{\tau'=0}^{\tau} d\tau' \int_{\chi'=0}^{\chi} G(\chi, \tau | \chi', \tau') g^*(\chi', \tau') \, d\chi'. \qquad (9)$$

The Green's function for ( 8 ) is

$$G(\chi, \tau | \chi', \tau') = 2 \sum_{m=1}^{\infty} \zeta_m e^{-(\beta_m^2 + C)(\tau - \tau')} \sin(\beta_m \chi) \sin(\beta_m \chi'), \qquad (10)$$

where $\zeta_m = \frac{\beta_m^2 + H_2^2}{\beta_m^2 + H_2^2 + H_2}$ and the eigenvalues are solutions to $\beta_m \cot(\beta_m) = -H_2$. Substituting the Green's function ( 10 ) and the dimensionless source term into ( 9 ) results in

$$\theta(\chi, \tau) = 2 \sum_{m=1}^{\infty} \zeta_m e^{-(\beta_m^2 + C)\tau} \sin(\beta_m \chi) \int_{\tau'=0}^{\tau} e^{(\beta_m^2 + C)\tau'} \int_{\chi'=0}^{1} \sin(\beta_m \chi') \frac{v}{2\sqrt{\tau'}} \delta(\chi' - v\sqrt{\tau'}) \, d\chi' \, d\tau'. \qquad (11)$$

The integral with respect to $\chi'$ in ( 11 ) is evaluated to

$$\int_{\chi'=0}^{1} \sin(\beta_m \chi') \frac{v}{2\sqrt{\tau'}} \delta(\chi' - v\sqrt{\tau'}) \, d\chi' = \frac{v}{2\sqrt{\tau'}} \sin(\beta_m v \sqrt{\tau'}) \qquad (12)$$

and the integral with respect to $\tau'$ in ( 11 ) is evaluated to

$$\int_{\tau'=0}^{\tau} e^{\varepsilon_m^2 \tau'} \frac{\sin(\beta_m v \sqrt{\tau'})}{\sqrt{\tau'}} d\tau' = e^{\left(\frac{\beta_m v}{2\varepsilon_m}\right)^2} \frac{\sqrt{\pi}}{\varepsilon_m} \left\{ \Re\left[\mathrm{erf}\left(\frac{\beta_m v}{2\varepsilon_m} + i\varepsilon_m \sqrt{\tau}\right)\right] - \mathrm{erf}\left(\frac{\beta_m v}{2\varepsilon_m}\right) \right\}, \qquad (13)$$

where $i$ is the imaginary unit, $\Re$ indicates the real component of the complex term, and $\varepsilon_m = \sqrt{\beta_m^2 + C}$. Substituting ( 12 ) and ( 13 ) into ( 11 ) leads to the dimensionless closed-form solution

$$\theta(\chi, \tau) = v\sqrt{\pi} \sum_{m=1}^{\infty} \zeta_m e^{-\varepsilon_m^2 \tau} \sin(\beta_m \chi) e^{\left(\frac{\beta_m v}{2\varepsilon_m}\right)^2} \frac{1}{\varepsilon_m} \left\{ \Re\left[\mathrm{erf}\left(\frac{\beta_m v}{2\varepsilon_m} + i\varepsilon_m \sqrt{\tau}\right)\right] - \mathrm{erf}\left(\frac{\beta_m v}{2\varepsilon_m}\right) \right\}. \qquad (14)$$

To evaluate this solution, the infinite sum is truncated at $N$, where $N$ is large enough for the sum to converge ($N = 50$ was found to be sufficient in most cases). A slower, more accurate, accelerated sum (using `nsum` in mpmath; Johansson et al., 2017) was used to check the faster truncated sum. For each evaluation of the sum, all eigenvalues $\beta_m$ ($m = 1, \ldots, N$) must be calculated. An iterative numerical algorithm was used to evaluate the transcendental eigenvalue equation (Haji-Sheikh & Beck, 2000).

Initially, the solution was developed using Laplace transforms and numerically inverted. The Green's function approach given here produces the same results as the Laplace transform approach, but it is faster and has fewer numerical issues at very early time. The solution is simple enough that it can readily be evaluated for parameter estimation.

**SOLUTION APPLICATION**

The dimensionless form of the governing equation includes three free parameters, namely $v$, $C$, and $H_2$. The reference temperature ($T_r$) is another free parameter which scales the solution by a constant factor when converting back to the dimensional form. These four parameters allow for the determination of the physical parameters, $k$, $\eta$, $h$, and $h_2$, given that $L$, $\sigma$, $\rho$, $c_p$ and $R$ (sample radius) are known.

A thermal imbibition experiment was conducted using a cylindrical core sample of tuff. The laboratory method is an extension of an approach used to estimate hydrologic properties from tuff by fitting numerical models to observed wetting front height and mass of water imbibed (Kuhlman et al., 2022b). The thermal imbibition experiment is presented in more detail (with photos) in Kuhlman et al. (2022a). The spontaneous imbibition experiment was performed on a zeolitic tuff sample from Aqueduct Mesa, Nevada (Prothro et al., 2009; Heath et al., 2021). The core has a diameter of 6.32 cm and a length of 10.4 to 11.4 cm (only one end of the sample is cut flat). The sample was dried in a convection oven at 60 °C for >72 hours before testing, when it was placed in a sample holder connected to a Mariotte bottle on a balance. The core had six resistance temperature detectors (RTDs) held to the circumference of the core with rubber bands. The RTDs were monitored with a National Instruments CompactDAQ using two analog data acquisition input modules (NI-9219 and NI-9217). Photographs were taken during the early phases of the test to document the progress of the wetting front. The sample holder was covered with a clear plastic bag, and the entire testing apparatus (sample holder, datalogger, and Mariotte bottle) was covered with a clear plastic box to minimize temperature fluctuations. Relative humidity within the outer box stayed around ~57% during the entire test.

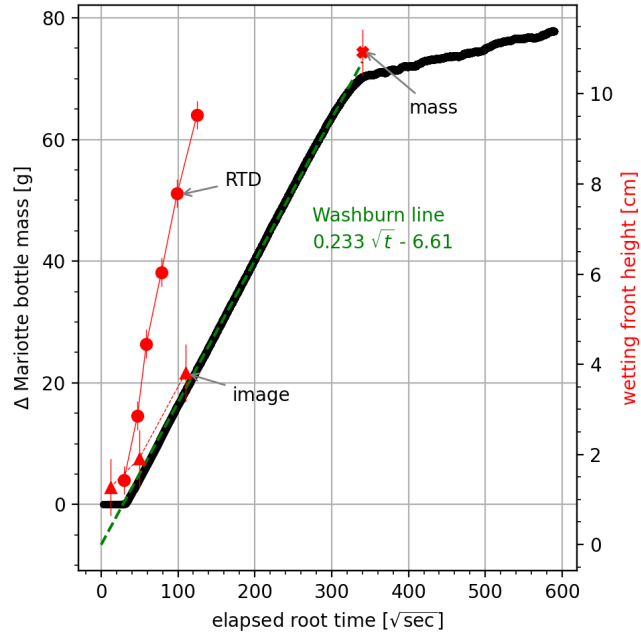

**Figure 2.** Mass imbibed data (black, left axis) and elevation data (red, right axis) for tuff. Wetting front elevation estimates from images (triangles) and RTDs (circles) show different sorptivities. Bend in mass imbibed data at 350 square-root seconds related to arrival of wetting front at top of sample (Kuhlman et al., 2022a), corresponding to red x labeled "mass". Green dashed line is best-fit Washburn (1921) straight line to early-time data.

**Mass Transport Model Parameter Estimation**

The sorptivity, $\sigma$, can be estimated separately by analyzing only the peaks in the temperature data. The primary temperature spike has been proposed to occur at the location of the wetting front (e.g., Prunty & Bell, 2005) or at the location of the vapor migration front (e.g., Anderson & Linville, 1960; Perrier & Prakash, 1977). Thus, the relation $s(t) = \sigma\sqrt{t}$ can be used to estimate $\sigma$ using the location and time of each peak listed in Table 1. Strong separation is observed between the temperature peak front (red circles) and the wetting front (from both images and the kink in the mass imbibed data) in Figure 2. This separation trend is analogous to, but stronger than that observed by Perrier & Prakash (1977). The kink in the mass imbibed data shows the wetting front reaches the top of the core much later (~25 hours to 9.5-cm RTD) than the thermal front (4.3 hours to top RTD). It can therefore be inferred for this zeolitic tuff, that the thermal pulse observed in the core is associated with adsorption of vapor diffusing away from the wetting front. The slow movement of the wetting front relative to the thermal front justifies the assumption of a static wetting front made in the mass transport model development.

The non-zero late-time slope of the mass imbibed data (Figure 2) is related to the bimodal pore-size distribution in the sample (Heath et al., 2021). The early slope is associated with the capillarity of the continuous fine-grained ash matrix, while the later slope is associated with isolated high-porosity pumice fragments with lower capillarity. This occurrence of two characteristic slopes is characteristic of a porous medium with two characteristic pore sizes (Ashraf et al., 2018). In this work we are concerned with the early time data, before the wetting front reaches the top of the sample.

Table 1. Times associated with peak temperature at each RTD sensor (red circles in Figure 2).

| Sensor | Sensor height (cm) | Peak temperature time (sec) | Time uncertainty (sec) |
|---|---|---|---|
| 9217 ai0 | 1.4 | 882 | 180 |
| 9217 ai1 | 2.9 | 2,238 | 180 |
| 9217 ai2 | 4.5 | 3,465 | 540 |
| 9217 ai3 | 6.5 | 6,133 | 1,440 |
| 9219 ai0 | 7.8 | 9,804 | 1,440 |
| 9219 ai1 | 9.5 | 15,561 | 2,160 |

Plots of the peak arrival data and the fitted model are shown in Figure 3. Using least-squares fitting estimates the vapor sorptivity, $\sigma_v = 7.15 \times 10^{-4}$ m/$\sqrt{s}$, which is then related to the effective vapor diffusion coefficient ($\sigma_v = 0.954\sqrt{\phi\,\bar{\tau}\,D_{wv,air}}$). Using $\phi = 0.225$ for the sample porosity (Kuhlman et al., 2022a) results in a tortuosity of $\bar{\tau} = 0.096$, which is somewhat small for a tuff—compare to tuff electrical tortuosities reported in the range 0.16 to 0.59 (Bernard et al., 2007). The porosity estimate is derived from imbibition testing and therefore may be low, due to effects of entrapped air. The actual diffusion of vapor in the sample likely deviates from the idealized solution, due to the adsorption of vapor, which delays the migration of the leading edge of the vapor front (like a retardation coefficient) and makes the tortuosity coefficient appear smaller (i.e., more tortuous). The presence of zeolite in the sample may also contribute to the smallness of the tortuosity, because of the relatively large fraction of nano-scale porosity and high specific surface area associated with zeolite. Estimation of tortuosity or more generally obstruction coefficients in tuffs is difficult (e.g., Paul et al., 2020), and this thermal monitoring method may be an effective method for constraining diffusive tortuosity in dry samples. More physically realistic numerical models for vapor migration in tuff, which account for vapor loss to adsorption, may improve the realism of the estimated tortuosity value.

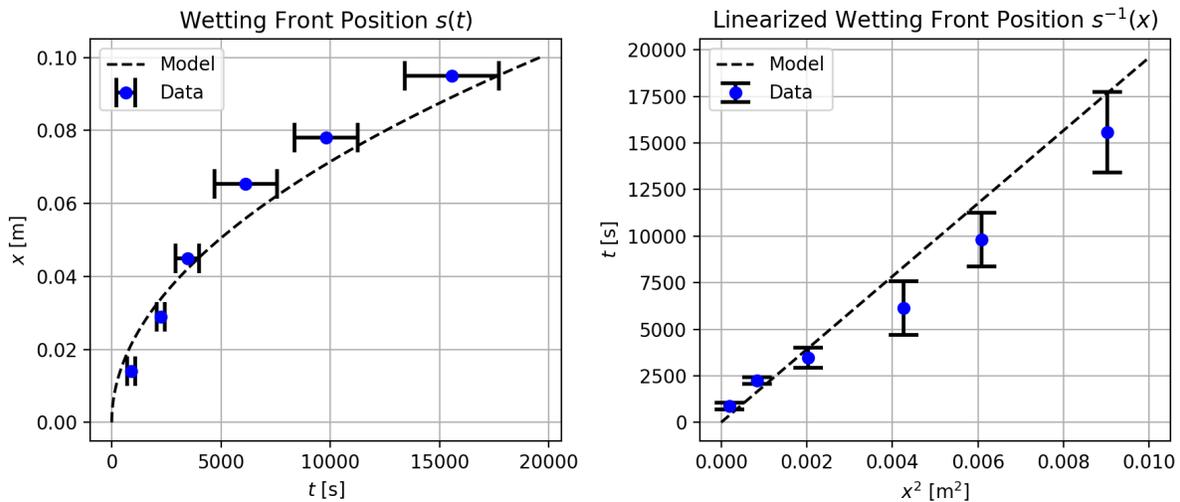

Figure 3. Time and location of thermal peaks observed via RTD.

**Energy Transport Model Parameter Estimation**

The Python SciPy optimize library (specifically `scipy.optimize.curve_fit`) was used to estimate the values of the free parameters in the thermal model evaluated using the Python mpmath library (Johansson et al., 2017), through fitting to the laboratory observations. The data were sampled uniformly, and each observation was weighted equally. Fitting the analytical solution to the data resulted in the estimates $T_r = 9.74 \pm 1.24$ K, $v = 0.935 \pm 0.074$, $C = 3.86 \pm 1.1$, and $H_2 = 3.31 \pm 0.72$. The uncertainties are 95% confidence intervals ($\pm 2$ standard deviations) from the square root of the covariance matrix diagonals. Given known values of $L = 0.1$ m, $\sigma_v = 7.15 \times 10^{-4}$ m/$\sqrt{s}$, $\rho = 1584$ kg/m$^3$, $c_p = 900$ J/(kg·K) (Hadgu et al., 2007), and $R = 0.03$ m, the unknown physical parameters can be calculated from the estimated dimensionless quantities. The physical parameters are thus $k = 0.83 \pm 0.13$ W/(m·K), $\eta = 8800 \pm 1120$ J/kg of bulk dry sample, $h = 4.8 \pm 1.4$ W/(m$^2$·K), and $h_2 = 28 \pm 7.4$ W/(m$^2$·K).

The analytical solution evaluated with the best-fit parameters is shown in Figure 4 with the original (unsampled) data. Observed change in temperature continued to decrease below the initial and boundary temperature at late time, due to differences in the temperature of the water and the tuff sample (likely due to evaporative cooling of the water reservoir), and due to fluctuations in the environment on the order of ~1 °C, despite the bag and clear plastic box over the test to reduce thermal fluctuations. Some of the difference in the peak temperature between the observations and the model can be explained by the finite width of the RTD (~1 cm), which averages the observed response over a short vertical interval, while the model predicts temperature at a point in space and time.

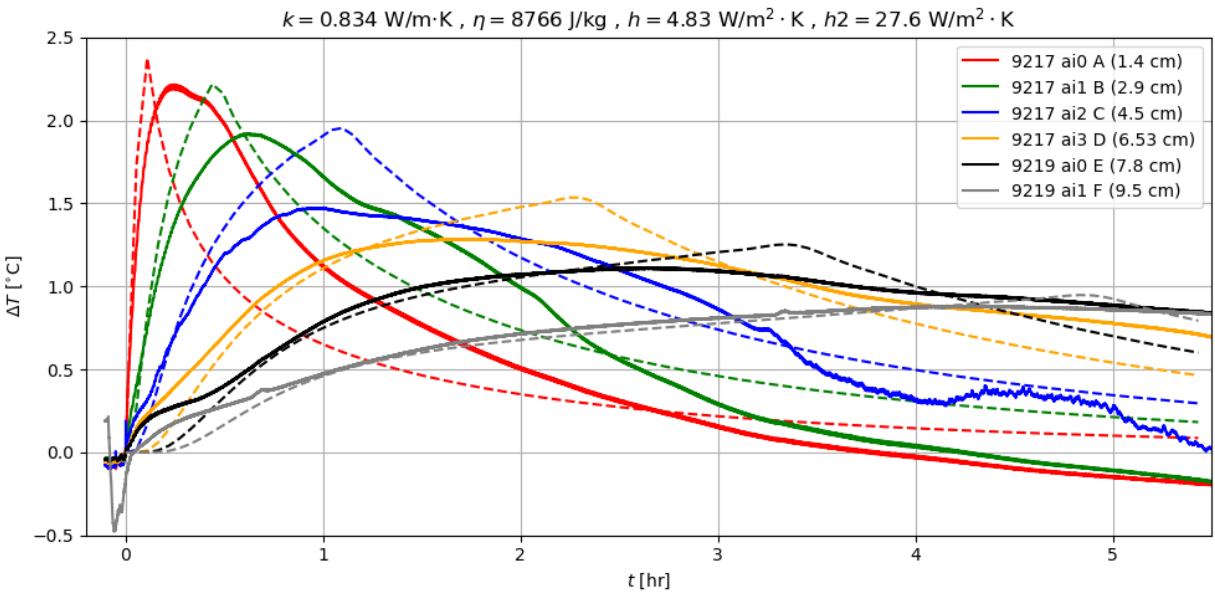

**Figure 4. Change in temperature ΔT above the initial background for both observations (solid) and model predictions (dashed) at RTD locations.**

This estimate of thermal conductivity is a bit lower, but generally consistent with estimates made in other tuffs (Johnstone & Wolfsberg, 1980; Sass et al., 1987; Barbero-Barrera et al., 2019). The estimate of $\eta$ from parameter estimation is lower than other values reported for energy released as part of the heat of wetting (Patrick & Grim, 1921; Anderson, 1924; Bright et al., 1953) and is more similar to heat of adsorption of vapor (Nordon & Bainbridge, 1983). If the heat loss coefficient from the sides or end of the sample were better known (i.e., possibly constrained from multiple tests with different types of insulation), the heat capacity of the sample could also be estimated independently. As the test was conducted, the only tuff thermal properties that could be estimated were the thermal conductivity and the heat source strength (the $C$ and $H_2$ Biot parameters are not physical properties of the rock, but properties of the core/laboratory testing system).

## SUMMARY

The laboratory experiment presented here clearly shows the thermal source in the tuff core is associated with vapor adsorption, since the thermal signal occurs hours before the wetting front is observed at the same location (either from extrapolation of early-time visual observations of the wetting front or the change in slope of mass imbibed data from the Mariotte bottle). Vapor is clearly migrating up the core, rather than condensing everywhere simultaneously due to vapor's high mobility (Aslannejad et al., 2017). This leads to a the relatively low effective vapor diffusion coefficient, retarding the vapor diffusion front compared to samples tested in other studies. The relative importance of vapor and water adsorption and the separation between the two depends strongly on sample characteristics (e.g., specific surface area and tortuosity). There is room for improvement in the fits between model and data (and in the possible design of future tests to better constrain thermal parameters), but the analytical approach presented here to estimate the sorptivity and thermal properties is simple to use and provides a clear advance in accurately predicting this phenomenon. This is a step towards developing a single laboratory test which can be used to simultaneously estimate single-phase flow (porosity and permeability), two-phase flow (van Genuchten (1980) model parameters), vapor transport (tortuosity), and thermal (thermal conductivity and source strength) properties. Kuhlman et al. (2022b) estimated the single- and two-phase flow parameters from mass imbibed and wetting front elevation data. The addition of the thermal data has not only increased the number of sample parameters estimable from the test (now including tortuosity, thermal conductivity, and source strength), but these observations have also added to the physical understanding of the processes that contribute to the imbibition of water into dry tuff (heat of wetting vs. heat of vapor adsorption).

The analytical solution was derived for a sorptivity that may be associated with a wetting front or a vapor front. The current approach is applicable to tests at small Péclet number (weak thermal advection and strong heat conduction), since no thermal advection term is included in the governing thermal equation. If the square of the sorptivity were much larger than the thermal diffusivity $\sigma^2/\alpha \gg 1$, advection of thermal energy with the water would be significant. The solution may be extended to include advection of heat, making the solution applicable to a wider range of samples. The physics of the system can also be implemented in numerical models, allowing advection of heat, non-linear behaviors, and arbitrary geometries or test configurations. Despite its simplicity, the analytical solution presented here does an excellent job of capturing the essential physical phenomena controlling the thermal signal observed during imbibition. This is the first use of the thermal imbibition response to estimate transport (tortuosity) and thermal properties (thermal conductivity and heat source strength) from a single laboratory test, which may be the most practical application of this phenomenon. Although multiple numerical models have implemented at least part of the physics related to heat of wetting and heat of adsorption (Milly & Eagleson, 1980; Benjamin et al., 1990; Prunty, 2002; Foss et al., 2003; Prunty & Bell, 2005; Murali et al., 2020), only one of these publications have plotted model predictions of temperature through time directly against observations made during imbition tests (Foss et al., 2003).

**ACKNOWLEDGMENTS**

This research was funded by the National Nuclear Security Administration, Defense Nuclear Nonproliferation Research and Development. The authors acknowledge important interdisciplinary collaboration with scientists and engineers from Los Alamos National Laboratory, Lawrence Livermore National Laboratory, Mission Support and Test Services, Pacific Northwest National Laboratory, and Sandia National Laboratories.

This article has been authored by an employee of National Technology & Engineering Solutions of Sandia, LLC under Contract No. DE-NA0003525 with the U.S. Department of Energy (DOE). The employee owns all right, title and interest in and to the article and is solely responsible for its contents. The United States Government retains and the publisher, by accepting the article for publication, acknowledges that the United States Government retains a non-exclusive, paid-up, irrevocable, world-wide license to publish or reproduce the published form of this article or allow others to do so, for United States Government purposes. The DOE will provide public access to these results of federally sponsored research in accordance with the DOE Public Access Plan https://www.energy.gov/downloads/doe-public-access-plan.

At Sandia National Laboratories, the authors thank Scott Broome for supervising the work, Jennifer Wilson for selecting the tuff core sample, and Tom Dewers for technically reviewing the manuscript.

**DECLARATION OF INTEREST**

The authors have no conflicts of interest to declare.